\documentclass[11pt]{article}

\def\mytitle#1{\setcounter{equation}{0}
\setcounter{footnote}{0}
\begin{flushleft}\Large\textbf{#1}\end{flushleft}
\vspace{0.25cm}}
\def\myname#1{\leftline{{\large #1}}\vspace{-0.13cm}}
\def\myplace#1#2{\small\begin{flushleft}\textit{#1}\\
\texttt{#2}\end{flushleft}}
\newenvironment{contribution}{\normalsize\noindent}{}
\def\myclassification#1{\small\noindent
Pacs no :
       #1\vspace{0.5cm}}
\usepackage{graphicx}% Include figure files
\begin{document}

\mytitle{Constraining Red-shift Parametrization Parameters in Brans-Dicke
Theory: Evolution of Open Confidence Contours}

\vskip0.2cm \myname{Ritabrata
Biswas\footnote{biswas.ritabrata@gmail.com}} \vskip0.2cm
\myname{Ujjal Debnath\footnote{ujjaldebnath@yahoo.com}}

\myplace{Department of Mathematics, Bengal Engineering and Science
University, Shibpur, Howrah-711 103, India.}{}

\begin{abstract}
In Brans Dicke theory of gravity, from the nature of the scalar field-potential considered, the dark energy, dark matter, radiation densities predicted by different observations and the closedness of the universe considered, we can fix our $\omega_{BD}$, the Brans Dicke parameter, keeping only the thing in mind that from different solar system constrains it must be greater than $5\times 10^{5}$. Once we have a value, satisfying the required lower boundary, in our hand we proceed for setting unknown parameters of the different dark energy models' EoS parameter. In this paper we work with three well known red shift parametrizations of dark energy EoS. To constrain their free parameters for Brans Dicke theory of gravity we take twelve point red shift vs Hubble's parameter data and perform $\chi^{2}$ test. We present the observational data analysis mechanism for Stern, Stern+BAO and Stern+BAO+CMB observations. Minimising $\chi^2$, we obtain the best fit values and draw different confidence contours. We analyze the contours physically. Also we examine the best fit of distance modulus for our theoretical models and the Supernovae Type  Ia Union2 sample. For Brans Dicke theory of gravity the difference from the mainstream confidence contouring method of data analysis id that the confidence contours evolved are not at all closed contours like a circle or a ellipse. Rather they are found to be open contours allowing the free parameters to float inside a infinite region of parameter space. However, negative EoSs are likely to evolve from the best fit values. 
\end{abstract}

\myclassification{04.60.Pp, 98.80.Qc}

\section{Introduction}

The simplest and best known one among scalar-tensor theories is
the Brans-Dicke (BD hereafter) theory of gravity \cite{Brans1}. Scalar Tensor theories of gravity include an extra scalar filed and hereby a potential dependent upon that besides the tensor part considered by the Einstein gravity.  Amongst them Brans-Dicke theory of gravity comprises with a constant parameter, named as BD parameter, which regulates the impact of the scalar fields inside the action as well as the field equations etc.The BD
theory, being a generalization of general relativity gives the
latter back with a high value of $\omega_{BD}$, the BD parameter
\cite{Barrow1}. Viking space probe says $\omega_{BD}$ should
exceed $500$ from timing experiments \cite{Reasenberg1}. The best
studied binaries with compact objects are the double neutron
stars, with the Hulse-Taylor pulsar $ (PSR 1913+16)$ as the
prototypical case. Unfortunately, in all double neutron-star
systems, the masses of the two members of the binary are
surprisingly similar and this severely limits the prospects of
placing strong constraints on the dipole radiation from them.
Indeed, the magnitude of dipole radiation depends on the
difference of the sensitivities between the two members of the
binaries, and for neutron stars the sensitivities depend primarily
on their masses. The resulting constraint imposed on the BD
parameter $\omega_{BD}$ by the Hulse-Taylor pulsar is
significantly smaller than the limit $\omega_{BD} > 40,000$ set by
the Cassini mission \cite{Bertotti1}. VLBI light deflection
theory predicts $\omega_{BD}$ to be $>3500$.\\

A massive scalar field has very negligible effect on the motion of
celestial bodies provided the mass is large enough with respect to
the inverse of the inter body distances. But if the mass is
sufficiently small, the corresponding potential $V(\phi)$ can be
locally neglected though the coupling function to matter will
strongly be constrained by experiment as we will see later on. In
case of solar system, the phenomena of precision is a good tool to
test from Newton's law to relativistic correction of it, namely the
post-Newtonian relativistic correction of it which is proportional
to $\frac{1}{c^2}$. Parametrized post-Newtonian correction
formalism is used to work with parameters of such orders. Two
famous parameters among them $\beta^{PPN}$ and $\gamma^{PPN}$
($\gamma=\frac{1+\omega}{2+\omega}$) introduced by Eddington
\cite{Eddington1, Will1} in the Schwarzchild metric [$-g_{00}
=1-\frac{2G m}{r c^2}+2\beta^{PPN}
\left(\frac{Gm}{rc^2}\right)^2+{\cal
O}\left(\frac{1}{c^6}\right)$;
$g_{ij}=\delta_{ij}\left(1+2\gamma^{PPN}
\frac{GM}{rc^2}\right)+{\cal O}\left(\frac{1}{c^4}\right)$] with
all the other eights are constrained to be very close to the
values of general relativity (in GR $\beta^{PPN}=\gamma^{PPN}=1$).
However, for scalar-tensor theories, the values are not unity any
more. The observed value of perihelion shift of mercury implies
the bound \cite{Shapiro1}
$$\left|2\gamma^{PPN}-\beta^{PPN}-1\right|<3\times 10^{-3}.$$
Lunar Laser Ranging \cite{Williams} gives the bound
$$\gamma^{PPN}-1=\left(2.1\pm 2.3\right)\times 10^{-5}.$$
L. Perivolaropoulos in \cite{Perivolaropoulos} has shown for
negligible mass of the field $\phi$, $\omega_{BD}$ having the
relation $\gamma=\frac{1+\omega}{2+\omega}$ turns to be
$\omega>4\times 10^4$ at the $2\sigma$ confidence level.
$\omega>5\times 10^4$ is supported value in some literature
\cite{Moon}. The light deflection as measured by Very Long
Baseline Interferometry \cite{Eubanks} gives the information
$$\left|\gamma^{PPN}-1\right|<4\times 10^{-4}.$$
The label ``Casini" to the impressive recent constraint obtained
by measuring the time delay variation to the Cassini spacecraft
near solar conjunction\cite{Bertotti1}:
$$\gamma^{PPN}-1=\left(2.1\pm 2.3\right)\times 10^{-5}.$$
So overall for solar system $\omega_{BD}>5\times 10^{4}$ will be
supported by all the different observation tools.\\

Dark Energy (DE), assumed to be distributed homogeneously all over
in the universe, is a component of the critical density of our
current universe as shown by the Cosmic Microwave Background (CMB)
and type Ia supernovae (SNIa) observations \cite{Riess1,
Perlmutter1, Spergel1, Knop1}. The cosmic standard candles, type
Ia SNe supernovae influence us to think about cosmic acceleration.
The Friedmann equation $\frac{\ddot{a}}{a}=-4\pi G
\left(\rho+3p\right)/3$ requires the condition
$\left(\rho+3p\right)<0$ for accelerated expansion ($\ddot{a}>0$).
As density is an ever positive physical quantity, we see the EoS
parameter must be negative and also less than $-1/3$. We give term
to such a negative pressure creating substance as DE. DE
occupies $73\%$ of the whole matter-energy of our universe.
Theoretically, we can find many proposed DE candidates. In
astrophysical sense it is popular to have a redshift
parametrization (i.e., taking the redshift $z$ as the variable
parameter of the EoS only) of the EoS as $p(z)=w(z) \rho(z)$. The
EoS parameter $w$ and its time derivative with respect to Hubble
time are currently constrained by the distance measurements of the
type Ia supernova and the current observational data constrain the
range of EoS as $-1.38<w<-0.82$ \cite{Melchiorri1}. Recently, the
combination of WMAP3 and Supernova Legacy Survey data shows a
significant constraint on the EoS $w=-0.97^{+0.07}_{-0.09}$ for
the DE, in a flat universe \cite{Seljak1}.\\

Two mainstream families of redshift parametrizations are there,
viz.,\\

$(i)$ Family $I$ : $w(z)=w_0 +w_1 \left(\frac{z}{1+z}\right)^n$
and\\

$(ii)$ Family $II$ : $w(z)=w_0 +w_1
\frac{z}{\left(1+z\right)^n}$,\\

where, $w_0$ and $w_1$ are two undecided parameters, $n$ is a
natural number. We will pick up three particular well known
parametrizations :

{\bf 1. Linear parametrization:} For $n=0$, family $II$ is known
as $``Linear~ parametrization"$
 $w(z)=w_{0}+w_{1}z$ \cite{Cooray1}. Here $w_{0}=-1/3$ and
$w_{1}=-0.9$ with $z<1$ when Einstein gravity has been considered.
This grows increasingly unsuitable for $z\gg 1$.
Upadhye-Ishak-Steinhardt parametrization \cite{Upadhye1} can avoid this
problem.

{\bf 2. CPL parametrization:} For $n=1$, both the families $I$ and
$II$ lead to the same parametrization
$w(z)=w_{0}+w_{1}\frac{z}{1+z}$. This ansatz was first discussed
by Chevallier and Polarski \cite{Chevallier1} and later studied
more elaborately by Linder \cite{Linder1}. In Einstein gravity the
best fit values for this model while fitting with the SN1a gold
data set are $w_{0} = -1.58$ and $w_{1} = 3.29$. This
parametrization will be shortly named as $``CPL~Parametrization"$
after the proposer Chevallier-Polarski-Lindler. There are
literature which supports that CPL parametrization has the
quantity to catch the dynamics of many DE models and in particular
the dynamics of the step like ones \cite{Linden1}.

{\bf 3. JBP parametrization:} For family $II$, $n=2$ gives the
parametrization $w(z)=w_{0}+w_{1}\frac{z}{(1+z)^{2}}$. A fairly
rapid evolution of this EoS allowed so that $w(z)\ge -1/2$ at
$z>0.5$ is consistent with the supernovae observation in Einstein
gravity. We will call this parametrization as $``JBP"$
\cite{Jassal1} parametrization.\\

Study of DE with different EoS value was studied in BD cosmology
by several authors. In 5D BD cosmology \cite{Errahmani1}, the
authors have shown that the DE component of the universe agrees
with the observational data. In \cite{Kim1} considering the
holographic energy density as a dynamical cosmological constant in
BD theory different future horizon cut-offs are been studied. The
work \cite{Arik1} tells the dependence of $H$ with a far different
cases of linearized non-vacuum solution in BD cosmology. They
predicted that the BD scalar field $\phi$ can explain DE but not
able to say about DM. The result of the study \cite{Kim2}
indicates that the BD scalar field appears to interpolate smoothly
between two late-time stages by speeding up the expansion rate of
the matter-dominated era somewhat while slowing down that of the
accelerating phase to some degree.\\

Wu and Chen \cite{Wu1} derived observational constraint on the BD
model in a flat FLRW universe with cosmological constant and cold
dark matter. For cosmic microwave back ground they had used , they
did include the WMAP five year data etc. They found degeneracy for
$\omega_{BD}<0$ for few data sets. In \cite{Li1} the authors used newly
published Planck CMB temperature data. The cosmological parameters
$H_0$, $\omega_{BDc} h^2$, $\sigma_8$ etc have been constrained.
Fabris et al \cite{Fabris} have studied cosmological solutions for
a pressureless fluid in the Brans-Dicke theory exhibit
asymptotical accelerated phase for some range of values of the
parameter $\omega_{BD}$, interpolating a matter dominated phase and an
inflationary phase. The effective gravitational coupling is
negative. The author did test this model against the supernovae
type Ia data. The fitting of the observational data is slightly
better than that obtained from the $\Lambda$CDM model. Finally
some speculations on how to reconcile the negative gravitational
coupling in large scale with the local tests, were made.\\

Our motive for this paper is to study the DE characterized by the
redshift parametrization of EoS in BD theory. We will try to
constrain the EoS parameters for different data sets. In the next
section (\ref{Basic Calculations}) we will construct the equations
for BD theory and the concerned parameters' expressions. to achieve the value of $\omega_{BD}$ which will be consistent with the solar system constrains we have calculated different $\omega_{BD}$s with respect to different $\alpha$ and chosen an appropriate one. In
section (\ref{data}) we will examine the best fitting values and
different sigma contours of $w_1$ and $w_2$ for Stern Data,
Stern+BAO and Stern+BAO+CMB respectively. Our main motive is to find the best fit values for $\omega_{0}$ and $\omega_{1}$. We will tally the
theoretical bound with the supernova data in the section
(\ref{Redshift-Magnitude Observations from Supernovae Type Ia}).
Finally, we will go for a brief summary in section
(\ref{Summary}).
%%%%%%%%%%%%%%%%%%%%%%%%%%%%%%%%%%%%%%%%%%%%%%%%%%%%%%%%%%%%%%%%%%%%%%%%%%%%%%
\section{Basic Equations for Brans-Dicke Theory}\label{Basic Calculations}
%%%%%%%%%%%%%%%%%%%%%%%%%%%%%%%%%%%%%%%%%%%%%%%%%%%%%%%%%%%%%%%%%%%%%%%%%%%%%%
The action of the self interacting Brans-Dicke theory reads as
\cite{Brans1} (choosing $c=1$)
\begin{equation}\label{BDaction}
S=\int \frac{d^4 x \sqrt{g}}{16\pi}\left[\phi R
-\frac{\omega_{BD}}{\phi} \phi^{,\alpha}\phi_{,\alpha}-V(\phi) + 16\pi
{\cal L}_m\right]
\end{equation}
here $\phi$ is the BD scalar field, $\omega_{BD}$ is the BD parameter,
$V(\phi)$ is the self interacting potential. In BD theory
$\frac{1}{\phi}$ exactly resembles with the factor $G$, the
gravitational constant. The action (\ref{BDaction}) also does
match with the low energy string theory action \cite{Sen1} for
$\omega_{BD}=-1$. The matter content of the universe is composed of DM,
DE and the radiation contribution.

From the Lagrangian density (\ref{BDaction}) we obtain the field
equation \cite{Sen1}
\begin{equation}\label{BDfield}
G_{\mu \nu} =\frac{8\pi}{\phi} T_{\mu \nu}^{m} +
\frac{\omega_{BD}}{\phi^2}
\left[\phi_{,\mu}\phi_{,\nu}-\frac{1}{2}g_{\mu \nu}
\phi_{,\alpha}\phi^{,\alpha}\right]+\frac{1}{\phi}\left[\phi_{,\mu
;\nu}-g_{\mu \nu}~ ^{\fbox{}}~ \phi \right]
-\frac{V(\phi)}{2\phi}g_{\mu \nu}
\end{equation}
and
\begin{equation}\label{BDphifield}
 ^{\fbox{}}~ \phi = \frac{8\pi
T}{3+2\omega_{BD}}-\frac{1}{3+2\omega_{BD}}\left[2V(\phi)
-\phi\frac{dV(\phi)}{d\phi}\right]
\end{equation}
where $T=T^{m}_{\mu\nu}g^{\mu\nu}$. Here,
$T_{\mu\nu}^{m}=(\rho+p)u_{\mu}u_{\nu}+p g_{\mu\nu}$ with
4-velocities $u^{\mu}$ obeying $u_{\mu}u^{\mu}=-1$.

Now choosing the line element for Friedmann-Robertson-Walker (FRW)
space time given by
\begin{equation}\label{FRW_Metric}
ds^2=-dt^2 +a^2(t)\left[\frac{dr^2}{1-kr^2}+r^2\left(d\theta^2 + sin^2 \theta d\phi^2 \right)\right]~~~~,
\end{equation}
where $a(t)$ is the scale factor and $k~(=0,~-1,~+1)$ is the
curvature index describe the flat, open and closed model of the
universe.

The Einstein field equations and the wave equation for the BD
scalar field for constant $\omega$ are given in the following \cite{Sen1}
\begin{equation}\label{Field_Eq_1}
H^2+\frac{k}{a^2}=\frac{8\pi
\rho_{tot}}{3\phi}-H\frac{\dot{\phi}}{\phi}+\frac{\omega_{BD}}{6}
\frac{\dot{\phi}^2}{\phi^2}+\frac{V(\phi)}{6\phi}~~~~and
\end{equation}
\begin{equation}\label{Field_Eq_2}
2\dot{H}+3H^{2}+\frac{k}{a^2}=-\frac{8\pi
p_{tot}}{\phi}-\frac{\omega_{BD}}{2}\frac{\dot{\phi}^2}{\phi^2}
-2H\frac{\dot{\phi}}{\phi}-\frac{\ddot{\phi}}{\phi}+\frac{V(\phi)}{2\phi}
\end{equation}
where, $H=\frac{\dot{a}}{a}$ is the Hubble parameter. If it is
assumed that matter is concerned in BD theory the conservation
equation is stated to be
\begin{equation}\label{continuity}
\dot{\rho}_{tot}+3H(\rho_{tot}+p_{tot})=0
\end{equation}
Now, our $\rho_{tot}$ comprises of densities of dark matter (DM),
dark energy (DE) and the density related to the radiation (i.e.,
$\rho_{tot}=\rho_{DM}+\rho_{DE}+\rho_{rad}$). Now, the pressure
corresponding these three components are respectively zero,
$p_{DE}$ governed by the EoS of concerned DE considered and one
third of the radiation density respectively. These immediately
gives us the total densities of the concerned fluids as
\begin{equation}\label{linear_density}
\rho_{tot}^{Linear}=\rho_{rad0}(1+z)^4+\rho_{DM0}(1+z)^3+\rho_{DE0}^{Linear}
\left(1+z\right)^{3(1+w_0-w_1)}\exp\left\{3w_1 z\right\}.
\end{equation}
and the same for CPL and JBP parametrization will be
\begin{equation}\label{CPL_density}
\rho_{tot}^{CPL}=\rho_{rad0}(1+z)^4+\rho_{DM0}(1+z)^3+\rho_{DE0}^{CPL}
\left(1+z\right)^{3(1+w_0+w_1)}\exp\left\{\frac{-3w_1 z}{1+z}\right\} ~~and
\end{equation}
\begin{equation}\label{JBP_density}
\rho_{tot}^{JBP}=\rho_{rad0}(1+z)^4+\rho_{DM0}(1+z)^3+\rho_{DE0}^{JBP}
\left(1+z\right)^{3(1+w_0)}\exp\left\{\frac{3w_1 z^2}{2(1+z)^2}\right\}~~ respectively.
\end{equation}
For simplicity of the calculation, we assume that
$V=V_{0}\phi^{n}$ and $\phi=\phi_{0}a^{\alpha}$. To determine $H$,
using equation (5) with equations (8) - (10), we have the
following expressions:
$$For~ Linear~ Parametrization: ~H^2+\frac{\alpha}{H_0}(1+z)HH_0
+\left[\left(\frac{k}{H_0^2}-\frac{\omega_{BD}}{6}\frac{\alpha^2}{H_0^2}\right)
(1+z)^2-\frac{V_0}{6H_0^2}
\phi_0^{n-1}\frac{1}{(1+z)^{\alpha(n-1)}}\right.$$
\begin{equation}\label{H_equation_Linear}\left.-\frac{8\pi}{3H_0^2}
\frac{1}{\phi_0}(1+z)^\alpha\left\{\rho_{rad0}(1+z)^4+\rho_{DM0}
(1+z)^3+\rho_{DE0}^{Linear}\left(1+z\right)^{3(1+w_0-w_1)}\exp\left\{3w_1 z\right\}\right\}\right]H_0^2=0
\end{equation}
$$For~ CPL~ Parametrization: ~H^2+\frac{\alpha}{H_0}(1+z)HH_0+
\left[\left(\frac{k}{H_0^2}-\frac{\omega_{BD}}{6}\frac{\alpha^2}
{H_0^2}\right)(1+z)^2-\frac{V_0}{6H_0^2}
\phi_0^{n-1}\frac{1}{(1+z)^{\alpha(n-1)}}\right.$$
\begin{equation}\label{H_equation_CPL}\left.-\frac{8\pi}{3H_0^2}
\frac{1}{\phi_0}(1+z)^\alpha\left\{\rho_{rad0}(1+z)^4+\rho_{DM0}
(1+z)^3+\rho_{DE0}^{CPL}\left(1+z\right)^{3(1+w_0+w_1)}\exp\left\{\frac{-3w_1 z}{1+z}\right\}\right\}\right]H_0^2=0
\end{equation}
$$For~ JBP~ Parametrization: ~H^2+\frac{\alpha}{H_0}(1+z)HH_0+
\left[\left(\frac{k}{H_0^2}-\frac{\omega_{BD}}{6}\frac{\alpha^2}
{H_0^2}\right)(1+z)^2-\frac{V_0}{6H_0^2}
\phi_0^{n-1}\frac{1}{(1+z)^{\alpha(n-1)}}\right.$$
\begin{equation}\label{H_equation_JBP}\left.-\frac{8\pi}{3H_0^2}
\frac{1}{\phi_0}(1+z)^\alpha\left\{\rho_{rad0}(1+z)^4+\rho_{DM0}
(1+z)^3+\rho_{DE0}^{JBP}\left(1+z\right)^{3(1+w_0)}\exp
\left\{\frac{3w_1 z^2}{2(1+z)^2}\right\}\right\}\right]H_0^2=0
\end{equation}
Defining the new parameters $\Omega_{\alpha 0}=\frac{\alpha
}{H_0}$, $\Omega_{k0}=\frac{k}{H_0^2}$, $\Omega_{V0}=\frac{V_0
}{6H_0^2}$, $\Omega_{i0}=\frac{8\pi}{3H_0^2}\rho_{i0}$ and
$E=H/H_0$ we have,
$$For~Linear ~Parametrizations~:~
E_{Linear}^2+\Omega_{\alpha
0}(1+z)E_{Linear}+\left[\left(\Omega_{k0}-\frac{\omega_{BD}}{6}\Omega_{\alpha
0}^2\right)(1+z)^2-\Omega_{V0}\phi^{n-1}\frac{1}{(1+z)^{\alpha
(n+1)}}\right.$$
\begin{equation}\label{E_Linear}
\left.-\frac{1}{\phi_0}\left\{\Omega_{rad 0}(1+z)^{4+\alpha}+\Omega_{DM0}
(1+z)^{3+\alpha}+\Omega_{DE0}^{linear}(1+z)^{\alpha+3(1+w_0-w_1)}exp\left\{
3w_1 z\right\}\right\}\right]=0
\end{equation}
$$For~CPL ~Parametrizations~:~
E_{CPL}^2+\Omega_{\alpha 0}(1+z)E_{CPL}+\left[\left(\Omega_{k0}-
\frac{\omega_{BD}}{6}\Omega_{\alpha 0}^2\right)(1+z)^2-\Omega_{V0}
\phi^{n-1}\frac{1}{(1+z)^{\alpha (n+1)}}\right.$$
\begin{equation}\label{E_CPL}
\left.-\frac{1}{\phi_0}\left\{\Omega_{rad 0}(1+z)^{4+\alpha}+\Omega_{DM0}
(1+z)^{3+\alpha}+\Omega_{DE0}^{CPL}(1+z)^{\alpha+3(1+w_0+w_1)}exp\left\{
\frac{-3w_1 z}{1+z}\right\}\right\}\right]=0
\end{equation}
$$For~JBP ~Parametrizations~:~
E_{JBP}^2+\Omega_{\alpha 0}(1+z)E_{JBP}+\left[\left(\Omega_{k0}-
\frac{\omega_{BD}}{6}\Omega_{\alpha 0}^2\right)(1+z)^2-\Omega_{V0}
\phi^{n-1}\frac{1}{(1+z)^{\alpha (n+1)}}\right.$$
\begin{equation}\label{E_JBP}
\left.-\frac{1}{\phi_0}\left\{\Omega_{rad 0}(1+z)^{4+\alpha}+\Omega_{DM0}
(1+z)^{3+\alpha}+\Omega_{DE0}^{JBP}(1+z)^{\alpha+3(1+w_0)}
\exp\left\{\frac{3w_1 z^2}{2(1+z)^2}\right\}\right\}\right]=0
\end{equation}
At the present universe $z=0$, so we get the condition
\begin{equation}\label{condition}
\left\{1+\Omega_{\alpha 0}+\Omega_{k0}-\frac{\omega_{BD}
}{6}\Omega_{\alpha 0}^2\right\}
\phi_0-\Omega_{V0}\phi_0^n-\left\{\Omega_{rad0}+\Omega_{DM0}+\Omega_{DE0}^{type}\right\}=0
\end{equation}
%We know, $\Omega_{\alpha 0}$, $\Omega_{k0}$, $\Omega_{V0}$,
%$\Omega_{rad0}$, $\Omega_{DM0}$ and $\Omega_{DE0}^{type}$ from
%physical observations. Here $type$ denotes for $Linear$, $CPL$ and
%$JBP$. So for our system we can find out the value of $\omega$. So
%far it can be speculated that $\omega$ depends on the total mass
%term
%$\Omega_{k0}-\left(\Omega_{rad0}+\Omega_{DM0}+\Omega_{DE0}^{type}\right)$.
%All other parameters concerned with the BD cosmology also affect
%it. Especially, it is inversely proportional to $\Omega_{\alpha
%0}$.
We fix some of the parameters using the best-fit values from
$7$ year WMAP data \cite{Komatsu1}. Now, $\Omega_{rad 0}=8.14
\times 10^{-5}$ and $\Omega_{DM0}=0.27$ and
$H_0=71.4Kmsec^{-1}/Mpc$ are the parametric values which are
determined from the physical observations (type denotes for
linear, CPL and JBP). We will take closed universe, i.e., $k=1$
which immediately will determine $\Omega_{k0}$. $\Omega_{V0}$ is
completely determined by the value of $V_0$ which will be taken as
a trivial value ($=1$). So for our model we have to determine
$\alpha$ (i.e., $\Omega_{\alpha 0}$) in such a way that satisfy
the observational constraint for $\omega_{BD}$. It is obvious that
$\alpha$ should be less than $1$, else the scalar field will
increase abruptly in late universe. For $\phi_0=V_0=1$, $n=2$
(i.e., $V(\phi)=\phi^{2}$) we have prepared the chart of
$\omega_{BD}$ vs $\alpha$ (given in Table 1). We follow that as
the value of $\alpha$ changes from $0.75$ to $0.7$, the value of
$\omega_{BD}$ exceeds $40,000$. From $7.0$ to $6.5$ it crosses
$50,000$. Now, beyond that as we decrease $\alpha$, $\omega_{BD}$
gets a high value. However we restrict ourself for $\alpha=0.5$
and proceed for the data analysis.
\[
\begin{tabular}{|c|c|c|}
\hline
  ~~~~~~$\alpha$ ~~~~& ~~~~$\omega_{BD}$~~~~\\
  \hline
  0.75 &  38734.6\\
  0.7 & 44421.7 \\

  0.68 & 47054.5 \\

  0.66 & 49929.6 \\
  0.65 & 51467.5 \\
  0.64 & 53077.9 \\

  0.62 & 56535.0 \\

  0.6 & 60342.9 \\
  0.5 & 86720.9 \\
   \hline
\end{tabular}
\]
{\bf Table 1:} Different values of $\omega_{BD}$ for different $\alpha$ chosen.\\

%%%%%%%%%%%%%%%%%%%%%%%%%%%%%%%%%%%%%%%%%%%%%%%%%%%%%%%%%%%%%%%%%%%%%%%%%%%%%%%%%%%%%%%%%%%%%%%%%%%%%%%%%%%%%%%%%%%%%%%%%%%%%%%%%%%%%
\section{Fitting with observational data}\label{data}
%%%%%%%%%%%%%%%%%%%%%%%%%%%%%%%%%%%%%%%%%%%%%%%%%%%%%%%%%%%%%%%%%%%%%%%%%%%%%%%%%%%%%%%%%%%%%%%%%%%%%%%%%%%%%%%%%%%%%%%%%%%%%%%%%%%%%%%%
Here, we are at the point to fit the observational data with our
model. Observed Hubble data at different redshifts (twelve data
points) given in observed Hubble data \cite{Stern} we will
proceed.  The Hubble parameter $H(z)$ and the standard error
$\sigma(z)$ for different values of redshift $z$ are given in
Table $2$. In the following subsections, we present the
observational data analysis mechanism for Stern, Stern+BAO and
Stern+BAO+CMB observations. We use the $\chi^{2}$ minimum test
from theoretical Hubble parameter with the observed data set and
find the best fit
values of unknown parameters for different confidence levels.\\

\[
\begin{tabular}{|c|c|c|}
\hline
  ~~~~~~$z$ ~~~~& ~~~~$H(z)$ ~~~~~& ~~~~$\sigma(z)$~~~~\\
  \hline
  0 & 73 & $\pm$ 8 \\
  0.1 & 69 & $\pm$ 12 \\
  0.17 & 83 & $\pm$ 8 \\
  0.27 & 77 & $\pm$ 14 \\
  0.4 & 95 & $\pm$ 17.4\\
  0.48& 90 & $\pm$ 60 \\
  0.88 & 97 & $\pm$ 40.4 \\
  0.9 & 117 & $\pm$ 23 \\
  1.3 & 168 & $\pm$ 17.4\\
  1.43 & 177 & $\pm$ 18.2 \\
  1.53 & 140 & $\pm$ 14\\
  1.75 & 202 & $\pm$ 40.4 \\ \hline
\end{tabular}
\]
{\bf Table 2:} The Hubble parameter $H(z)$ and the standard error
$\sigma(z)$ for different values of redshift $z$.\\

%%%%%%%%%%%%%%%%%%%%%%%%%%%%%%%%%%%%%%%%%%%%%%%%%%%%%%%%%%%%%%%%%%%%%%%%%%%%%%%%%%%%%%%%%%%%%%%%%%%%%%%%%%%%%%%%%%%%%%%%%%%%%%%%%%%%%
\subsection{Constraining Tool : $H(z)$-$z$ (Stern) data }\label{Analysis For
Stern Data}
%%%%%%%%%%%%%%%%%%%%%%%%%%%%%%%%%%%%%%%%%%%%%%%%%%%%%%%%%%%%%%%%%%%%%%%%%%%%%%%%%%%%%%%%%%%%%%%%%%%%%%%%%%%%%%%%%%%%%%%%%%%%%%%%%%%%%%%%
We first form the $\chi^{2}$ statistics as a sum of standard normal
distribution as follows: For any data set we will calculate the
minimum $\chi^{2}$, with the formula
\begin{equation}
{\chi}^{2}_{Stern}=\sum\frac{(H(z)-H_{obs}(z))^{2}}{\sigma^{2}(z)}
\end{equation}
Here, for different redshifts the theoretical and observational
values of Hubble parameter is given as $H(z)$ and $H_{obs}(z)$.
The corresponding error term is given as $\sigma(z)$. This is
however given in $Table~ 2$. In this statistics, the nuisance
parameter is given by $H_{obs}$ which can be safely marginalized.
Considering $H_0$ to have a fixed
prior distribution we will proceed.

This mechanism has recently been also discussed by several authors
\cite{Wu,Paul1,Paul2,Paul3,Paul4,Chak} in very simple way. Here we
shall determine the parameters $w_{0}$ and $w_{1}$ from minimizing
the above distribution ${\chi}^{2}$. The probability distribution
function in terms of the parameters $w_{0}$ and $w_{1}$ can be
written as

\begin{equation}
L= \int e^{-\frac{1}{2}{\chi}^{2}_{Stern}}P(H_{0})dH_{0}
\end{equation}

where $P(H_{0})$ is the prior distribution function for $H_{0}$.
We now plot the graph for different confidence levels (like 66\%,
90\% and 99\%).

\begin{figure}
~~~~~~~~~~~~~~~~~~~~~~Fig.1a~~~~~~~~~~~~~~~~~~~~~~~~~~~~~~~~~~~~~~~
Fig.1b~~~~~~~~~~~~~~~~~~~~~~~~~~~~~~~~~~~~~~~Fig.1c\\
\includegraphics[height=2in, width=2in]{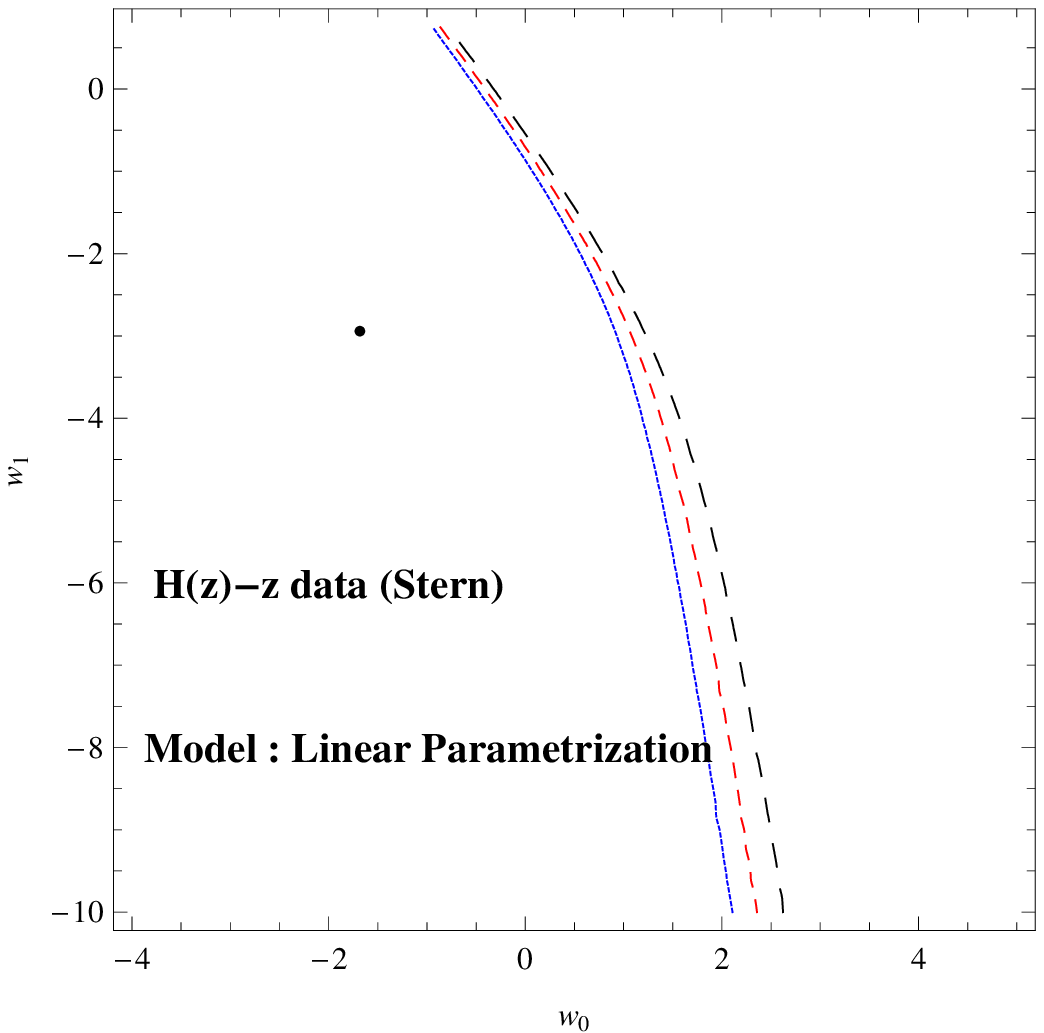}~~~~~~~
\includegraphics[height=2in, width=2in]{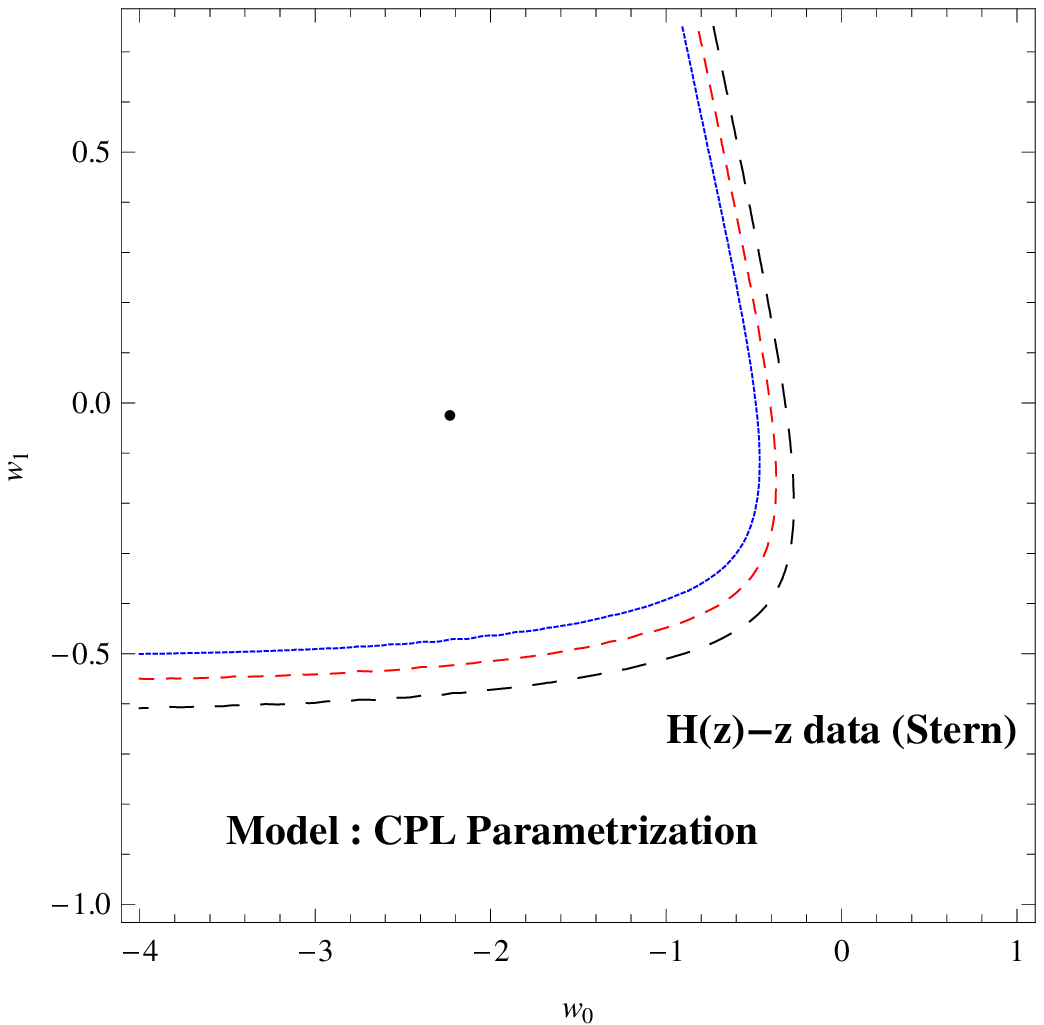}~~~~~~~
\includegraphics[height=2in, width=2in]{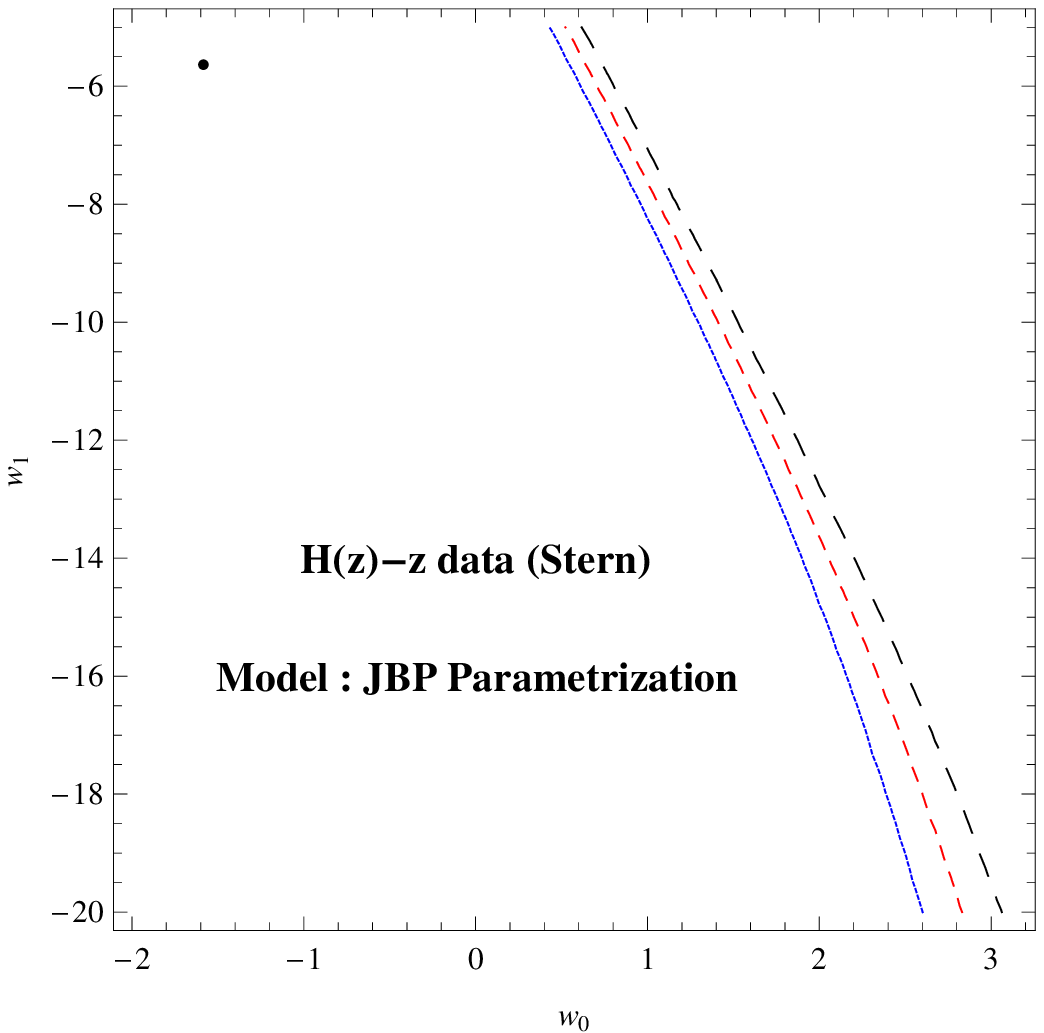}\\

\vspace{1cm}

Figs. 1(a), (b), (c) show that the variation of $w_{0}$ with
$w_{1}$ for different confidence levels. The 66\% (solid, blue,
the innermost contour), 90\% (dashed, red, next to the inner most
contour) and 99\% (dashed, black, the outermost contour) contours
are plotted in these figures for the $H(z)$-$z$ (Stern) analysis
(For Linear, CPL and JBP parameterizations respectively).
\vspace{1cm}
\end{figure}

The values of $w_0$ and $w_1$ (for which we can obtain the least
$\chi^2$) are given in the first row of the Table 3. We have
plotted the $1\sigma$, $2\sigma$ and $3\sigma$ confidence contours
in 1a-c curves. For BD cosmology one new aspect can be followed.
Here the contours are not closed. They are open curves. As for
example for linear parametrization the less $w_1$, less $w_0$
zones than the best fit $w_0$ and $w_1$ are included in the
$1\sigma$ curves. For CPL parametrization the less $w_0$, higher
$w_1$ area is included in the $1\sigma$ region. Lastly, for JBP we
can see if we choose a particular $w_0$ we need to get high $w_1$
to be in the $1\sigma$ zone. Similarly, for a fixed high $w_1$ it is
required to take low $w_0$ to be inside the $1\sigma$ contour. The
ultimate over all trend tells that we need sufficient negative
$w_0$ and $w_1$, which may evolve a negative  $w(z)$ ultimately.
Linear and CPL parametrizations are quite strict for this negative
$w(z)$ fact. But CPL however, allows positive $w(z)$ into
$1\sigma$ contour. Nevertheless it has its best fit in negative
$w(z)$ area.

%%%%%%%%%%%%%%%%%%%%%%%%%%%%%%%%%%%%%%%%%%%%%%%%%%%%%%%%%%%%%%%%%%%%%%%%%%%%%%%%%%%%%%%%%%%%%%%%%%
\subsection{Joint Analysis with Stern+BAO Data Sets}\label{Joint Analysis with Stern+BAO Data Sets}
%%%%%%%%%%%%%%%%%%%%%%%%%%%%%%%%%%%%%%%%%%%%%%%%%%%%%%%%%%%%%%%%%%%%%%%%%%%%%%%%%%%%%%%%%%%%%%%%%%

We will follow the pathway shown by Eisenstein et al
\cite{Einstein} for joint analysis, the Baryon Acoustic
Oscillation (BAO) peak parameter value. Here we will follow their
approach. Sloan Digital Sky Survey (SDSS) survey is one of the
first redshift survey by which the BAO signal has been directly
detected at a scale $\sim 100 ~MPc$. For low redshift ($0<z<0.35$)
we will check for the BAO peak to determine the DE parameters. The
BAO peak parameters might be defined as
\begin{equation}
{\cal A}=\frac{\sqrt{\Omega_m}}{E(z_1)^{\frac{1}{3}}}\left(\frac{1}{z_1}\int_{0}^{z_1}\frac{dz}{E(z)}\right)
^{\frac{2}{3}}
\end{equation}
Here $E(z)=H(z)/H_{0}$ is the normalized Hubble parameter, the
redshift $z_{1}=0.35$ is the typical redshift of the SDSS sample
and the integration term is the dimensionless comoving distance to
the redshift $z_{1}$ The value of the parameter ${\cal A}$ for the
flat model of the universe is given by ${\cal A}=0.469\pm 0.017$
using SDSS data \cite{Einstein} from luminous red galaxies survey.
Now the $\chi^{2}$ function for the BAO measurement can be written
as

\begin{equation}
\chi^{2}_{BAO}=\frac{({\cal A}-0.469)^{2}}{(0.017)^{2}}
\end{equation}

Now the total joint data analysis (Stern+BAO) for the $\chi^{2}$
function may be defined by

\begin{equation}
\chi^{2}_{total}=\chi^{2}_{Stern}+\chi^{2}_{BAO}
\end{equation}

According to our analysis the joint scheme gives the best fit
values of $w_{0}$ and $w_{1}$ in the second row of Table 3. Finally we
draw the contours $w_{0}$ vs $w_{1}$ for the 66\%
(solid, blue), 90\% (dashed, red) and 99\%
(dashed, black) confidence limits depicted in figures 2a to 2c.\\

\begin{center}
\begin{tabular}{|l|}
\hline\hline
Constraining Tool~~~Name of the Dark energy model~~~~~Best fit values of $w_0$, $w_1$ and $\chi^2$
\\\hline

\\
~~~~~~~~~~~~~~~~~~~~~~~~~~~~~~~~~~~~~~~~~~~~~~~~~~~~~~~~~~~~~~~~~~~~~~~~~$w_0$~~~~~~~~~~~~~~~~$w_1$~~~
~~~~~~~~~~~~~~$\chi^2$
\\ \hline \\
~~~~~~~~~~~~~~~~~~~~~~~~~~~~~~~~~~~~~~~~~~~Linear~~~~~~~~~~~~~~~~~~-1.68326~~~~~~~~~~-2.94136
~~~~~~~~7.30381

\\
Stern~~~~~~~~~~~~~~~~~~~~~~~~~~~~~~~~~~~~CPL~~~~~~~~~~~~~~~~~~~~-2.23223~~~~~~~~-0.0250015
~~~~~~~~7.32474

\\
~~~~~~~~~~~~~~~~~~~~~~~~~~~~~~~~~~~~~~~~~~~JBP~~~~~~~~~~~~~~~~~~~~-1.58429~~~~~~~~-5.6347
~~~~~~~~~~7.31521

\\\hline\hline\\

~~~~~~~~~~~~~~~~~~~~~~~~~~~~~~~~~~~~~~~~~~~Linear~~~~~~~~~~~~~~~~~-1.68687~~~~~~~~~-3.27207
~~~~~~~768.128

\\
Stern+BAO~~~~~~~~~~~~~~~~~~~~~~~~~~~~CPL~~~~~~~~~~~~~~~~~~~-2.26669~~~~~~~~~~-0.02538
~~~~~~~~~768.14821

\\
~~~~~~~~~~~~~~~~~~~~~~~~~~~~~~~~~~~~~~~~~~~JBP~~~~~~~~~~~~~~~~~~~~-1.70844~~~~~~~~~~-2.37717
~~~~~~~~~768.144

\\\hline\hline \\

~~~~~~~~~~~~~~~~~~~~~~~~~~~~~~~~~~~~~~~~~~~Linear~~~~~~~~~~~~~~~~~~-1.64718~~~~~~~~-4.35378
~~~~~~~~9962.81

\\
Stern+BAO+CMB~~~~~~~~~~~~~~~~~~CPL~~~~~~~~~~~~~~~~~~~~-2.52016~~~~~~~-0.0292668~~~~~~~~~9963.47

\\
~~~~~~~~~~~~~~~~~~~~~~~~~~~~~~~~~~~~~~~~~~~JBP~~~~~~~~~~~~~~~~~~~~-1.59007~~~~~~~~-6.00811
~~~~~~~~~~~9962.82

\\ \hline\hline
\end{tabular}
\end{center}
{\bf Table 3:} Best fit values of $w_0$, $w_1$ and $\chi^2$ for
Linear, CPL and JBP parametrizations models of dark energy in Stern,
Stern+BAO and Stern+BAO+CMB observations.\\

Like Stern data analysis, for Stern+ BAO also we get the open
contours while drawing the different $\sigma$ curves. For Liner
parametrization the contours are open downwards and a bit more oblique
towards the negative $w_0$ axis(if we compare with the Stern case). Which immediately tells us for this
case if $w_1$ is comparatively low (though positive!) we can vary
$w_0$ as we wish. The best fit value lie in the third quadrant.
 For CPL parametrization the scenario remains exactly the same. The best fit $w_0$ and $w_1$ are in the second quadrant
of the $\left(w_0, w_1\right)$ space. JBP parametrization also requires a
third-quadrant situated best fit for the minimum $\chi^2$. Here
$w_0$ is bounded at right like linear parametrization case. The sigma contours have a negative slope, i.e., as we
decrease our $w_0$ we can increase $w_1$ as well to stay inside
the $1\sigma$ contour.
\begin{figure}
~~~~~~~~~~~~~~~~~~~~~~~Fig.2a~~~~~~~~~~~~~~~~~~~~~~~~~~~~~~~~
~~~~~~~Fig.2b~~~~~~~~~~~~~~~~~~~~~~~~~~~~~~~~~~~~~~~Fig.2c\\
\includegraphics[height=2in, width=2in]{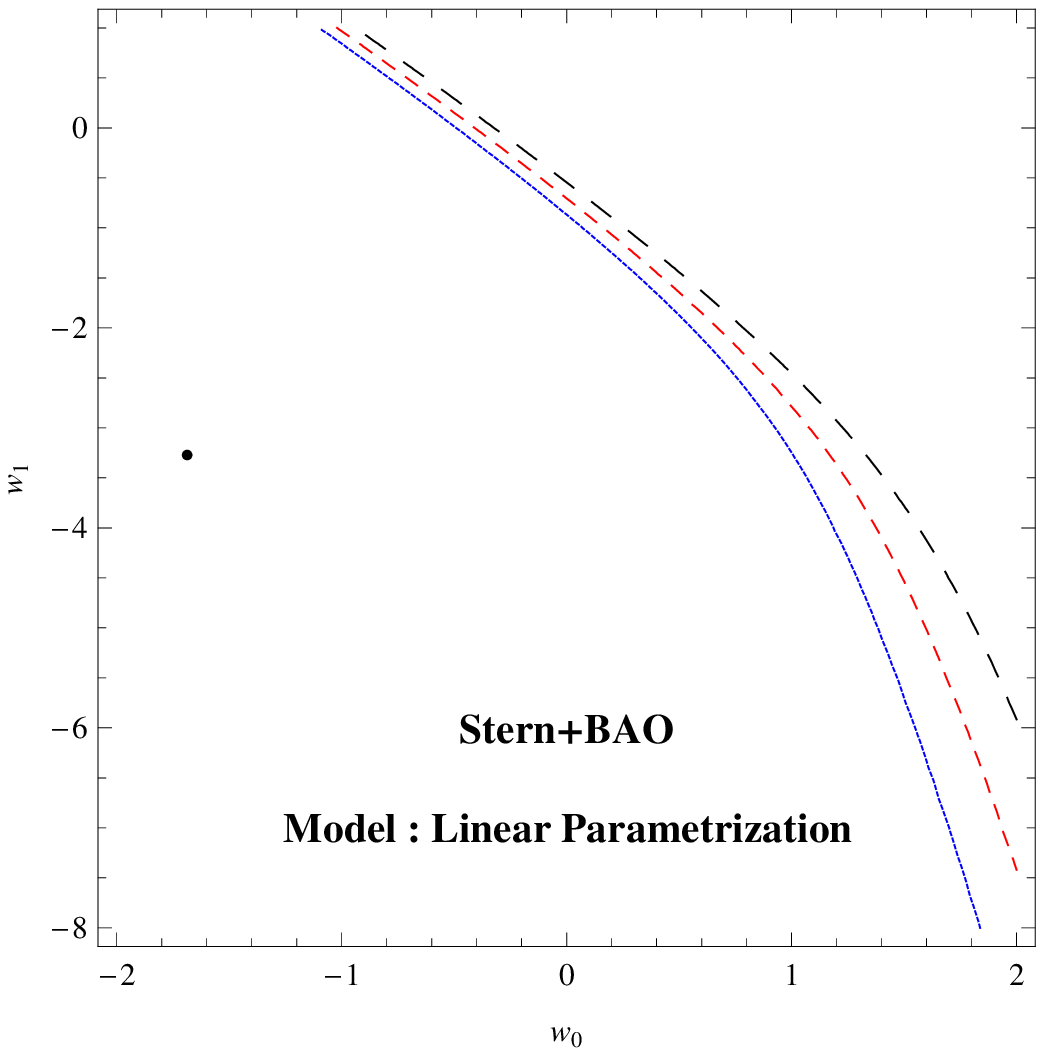}~~~~~~~
\includegraphics[height=2in, width=2in]{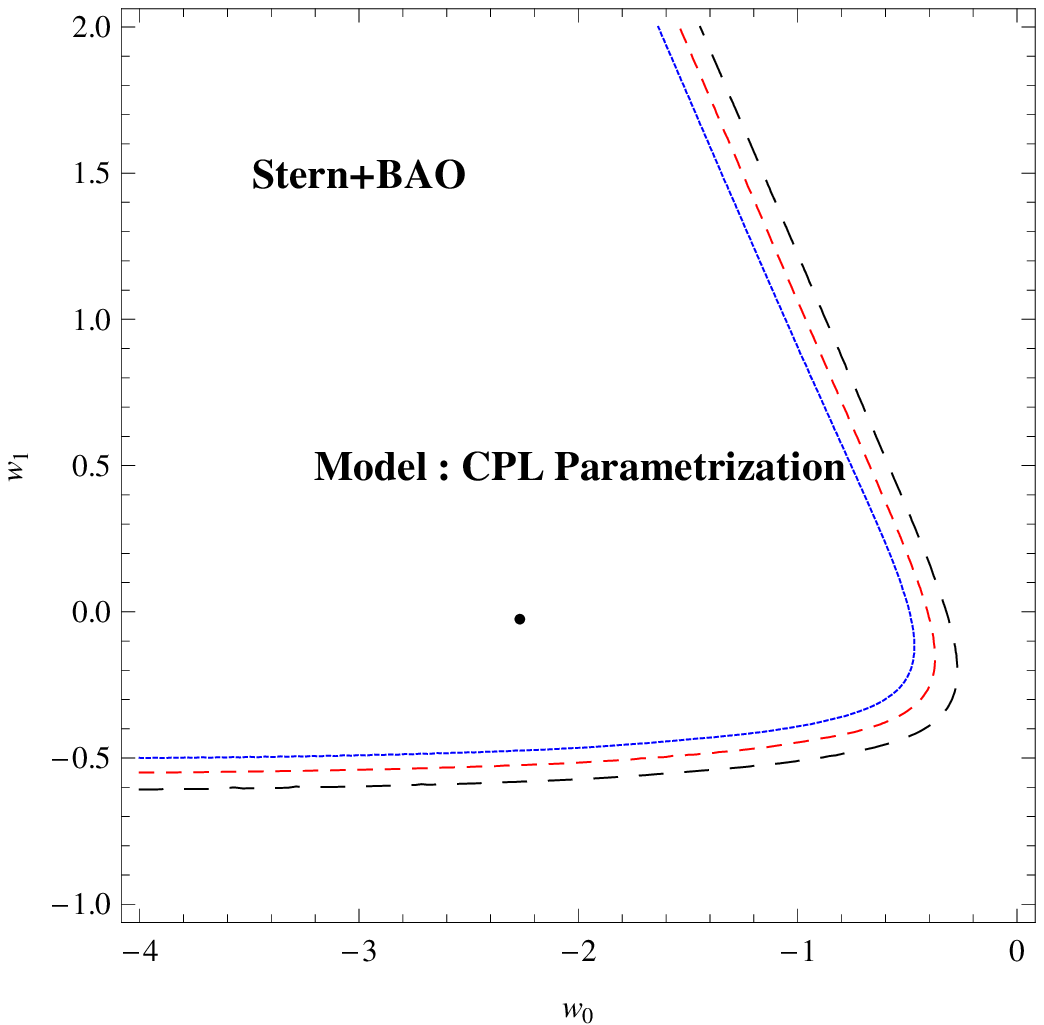}~~~~~~~
\includegraphics[height=2in, width=2in]{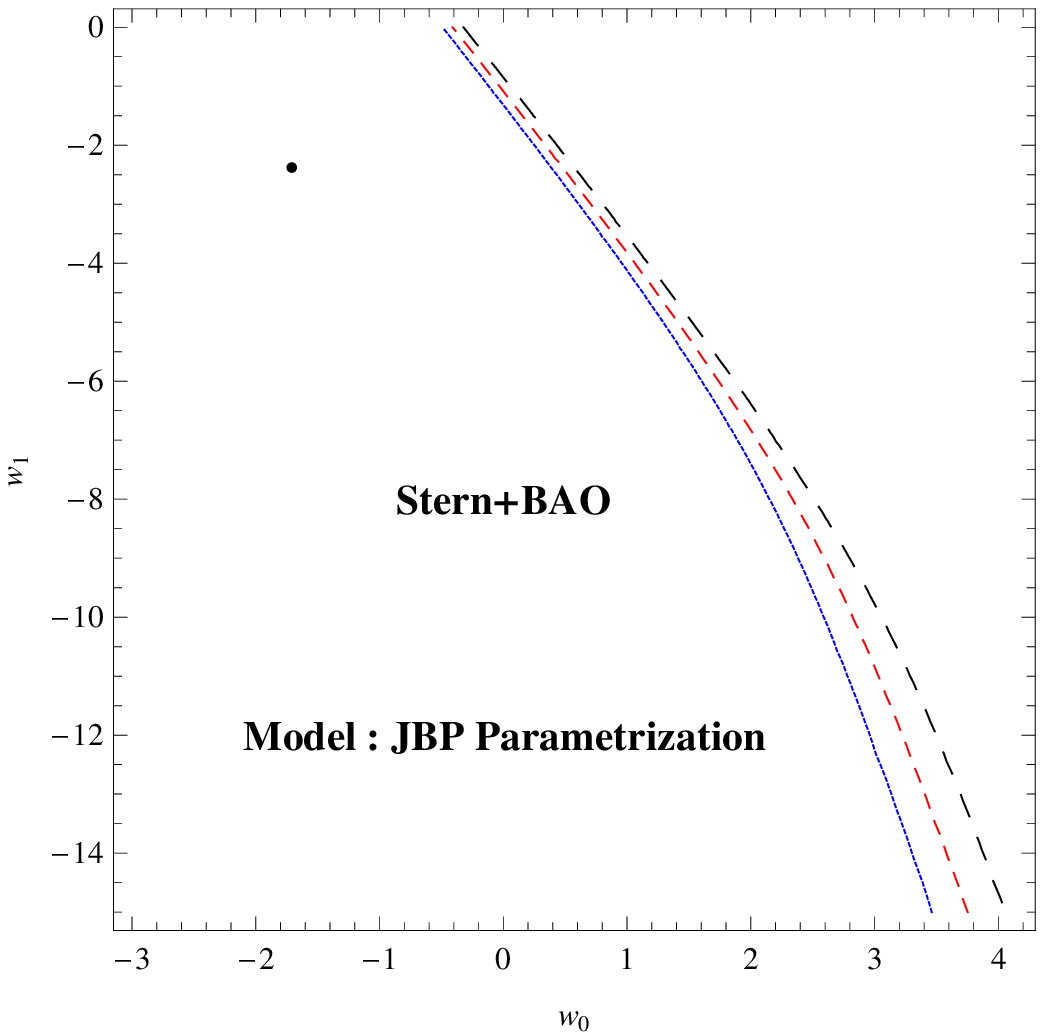}\\

\vspace{1cm}

Figs. 2(a), (b), (c) show that the variation of $w_{0}$ with
$w_{1}$ for different confidence levels. The 66\% (solid,
blue, the innermost contour), 90\% (dashed, red, next to the inner
most contour) and 99\% (dashed, black, the outermost contour)
contours are plotted in these figures for the $H(z)$-$z$
(Stern+BAO) analysis (For Linear, CPL and JBP parameterizations
respectively). \vspace{1cm}
\end{figure}

%%%%%%%%%%%%%%%%%%%%%%%%%%%%%%%%%%%%%%%%%%%%%%%%%%%%%%%%%%%%%%%%%%%%%%%%%%%%%%%%%%%%%%%%%%%%%%%%%
\subsection{Joint Analysis with Stern + BAO + CMB Data Sets}
\label{Joint Analysis with Stern + BAO + CMB Data Sets}
%%%%%%%%%%%%%%%%%%%%%%%%%%%%%%%%%%%%%%%%%%%%%%%%%%%%%%%%%%%%%%%%%%%%%%%%%%%%%%%%%%%%%%%%%%%%%%%%%%

In this subsection, we shall follow the pathway, proposed by some
author \cite{Bond,Efstathiou,Nessaeris}, using Cosmic Microwave
Background (CMB) shift parameter. One interesting geometrical
probe of DE can be determined by the angular scale of the first
acoustic peak through angular scale of the sound horizon at the
surface of last scattering which is encoded in the CMB power
spectrum. It is not sensitive with respect to perturbations but
are suitable to constrain model parameter. The CMB power spectrum
first peak is the shift parameter which is given by

\begin{equation}
{\cal R}=\sqrt{\Omega_{m}} \int_{0}^{z_{2}} \frac{dz}{E(z)}
\end{equation}
where $z_{2}$ is the value of redshift at the last scattering
surface. From WMAP7 data of the work of Komatsu et al
\cite{Komatsu1} the value of the parameter has obtained as ${\cal
R}=1.726\pm 0.018$ at the redshift $z=1091.3$. Now the $\chi^{2}$
function for the CMB measurement can be written as

\begin{equation}
\chi^{2}_{CMB}=\frac{({\cal R}-1.726)^{2}}{(0.018)^{2}}
\end{equation}

Now when we consider three cosmological tests together, the total
joint data analysis (Stern+BAO+CMB) for the $\chi^{2}$ function
may be defined by

\begin{equation}
\chi^{2}_{TOTAL}=\chi^{2}_{Stern}+\chi^{2}_{BAO}+\chi^{2}_{CMB}
\end{equation}
Now the best fit values of $w_0$ and $w_1$ for joint analysis of BAO
and CMB with Stern observational data support the theoretical
range of the parameters given in the third row of Table 3. The 66\% (solid, blue),
$90\%$ (dashed, red) and $99\%$ (dashed, black) contours are plotted
in figures $3a-3c$.

\begin{figure}
~~~~~~~~~~~~~~~~~~~~~~~Fig.3a~~~~~~~~~~~~~~~~~~~~~~~~~~~~~~~~~~~~~~~
Fig.3b~~~~~~~~~~~~~~~~~~~~~~~~~~~~~~~~~~~~~~~Fig.3c\\
\includegraphics[height=2in, width=2in]{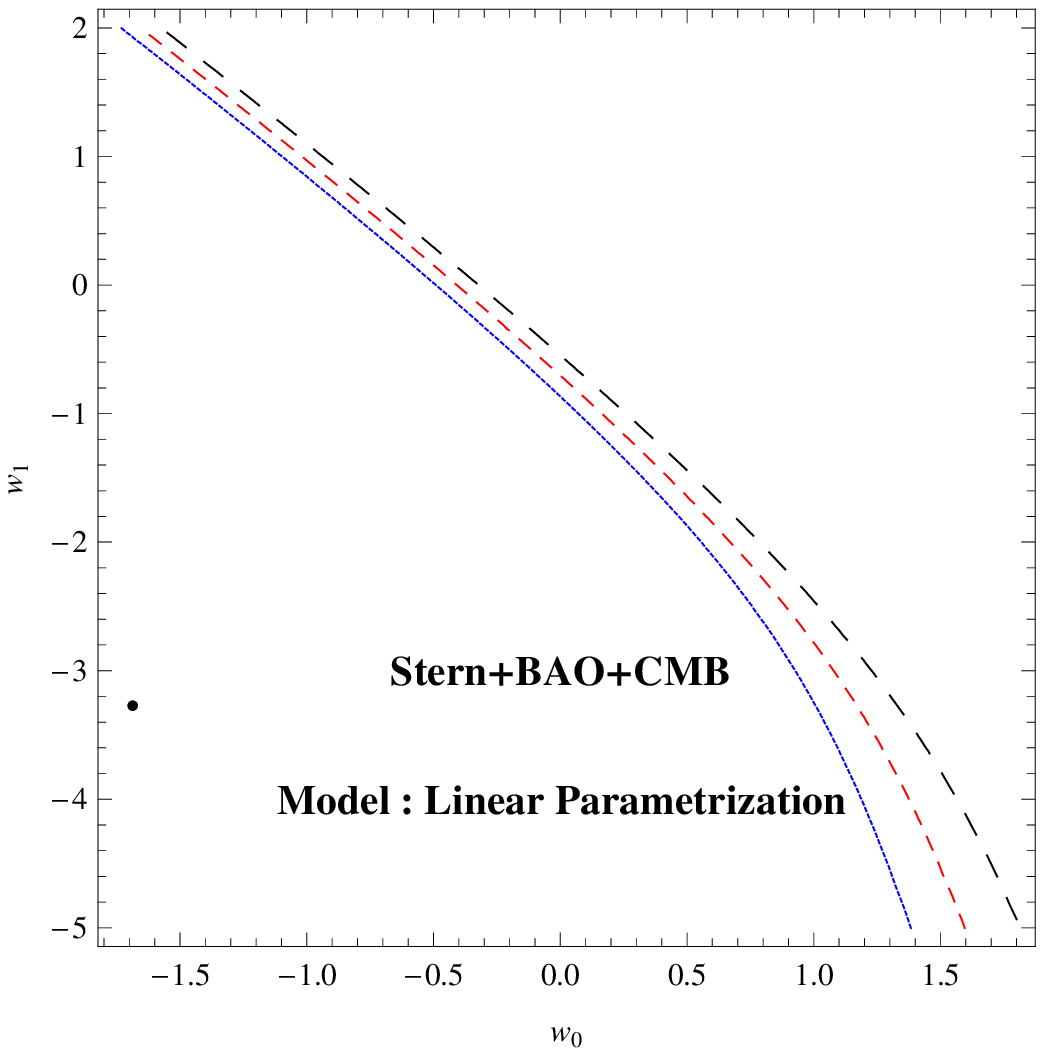}~~~~~~~
\includegraphics[height=2in, width=2in]{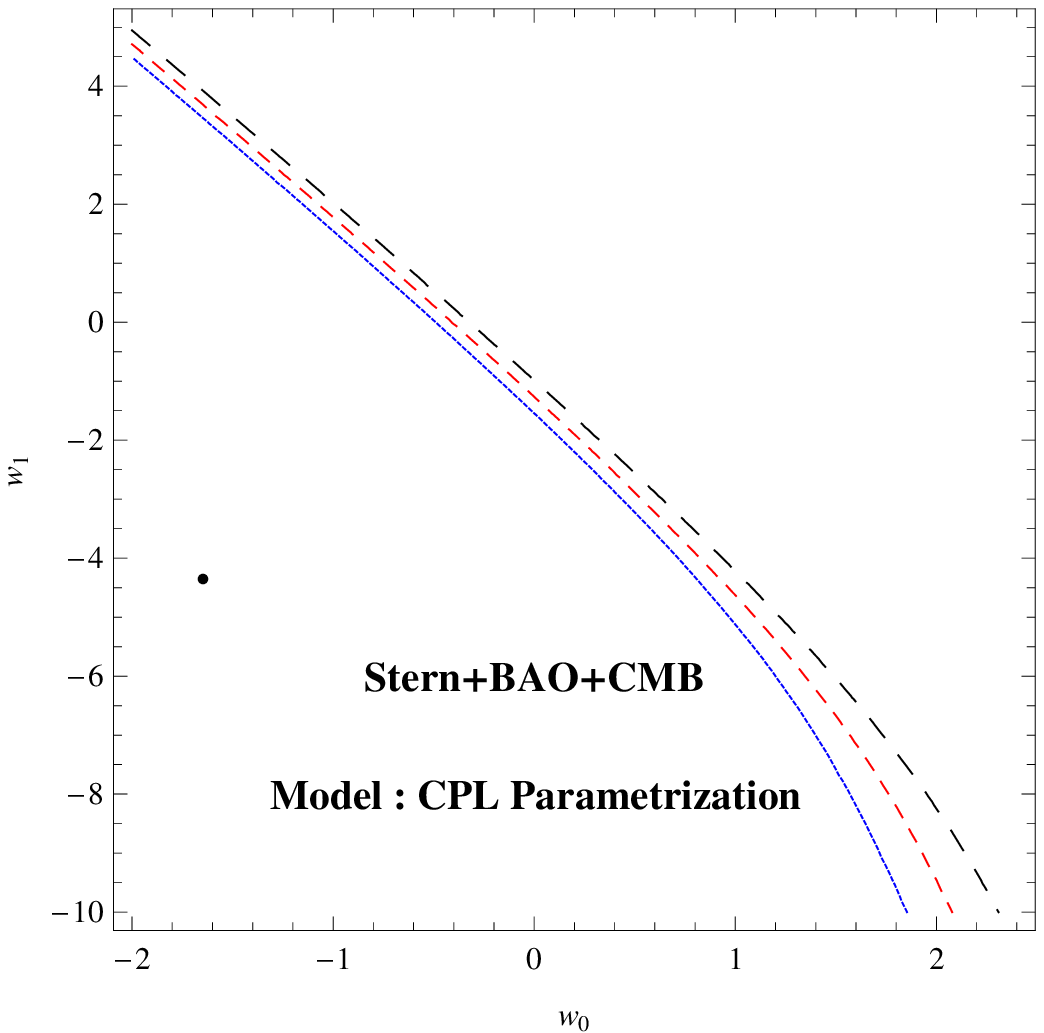}~~~~~~~
\includegraphics[height=2in, width=2in]{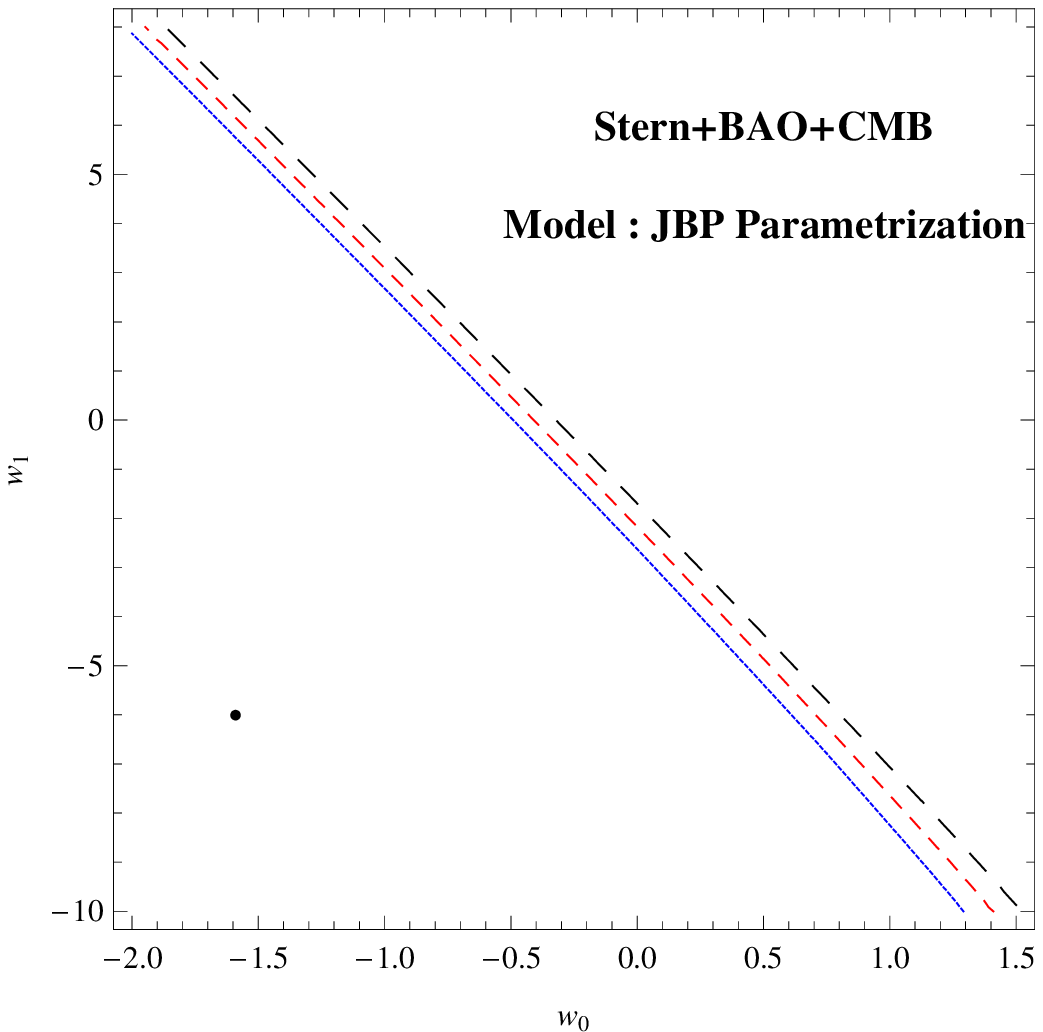}\\

\vspace{1cm}

Figs. 3(a), (b), (c) show that the variation of $w_{0}$ with
$w_{1}$ for different confidence levels. The 66\% (solid, blue,
the innermost contour), 90\% (dashed, red, next to the inner most
contour) and 99\% (dashed, black, the outermost contour) contours
are plotted in these figures for the $H(z)$-$z$ (Stern+BAO+CMB)
analysis (For Linear, CPL and JBP parameterizations respectively).
\vspace{1cm}
\end{figure}

At a glance, the sigma contours for Stern+BAO+CMB analysis
resemble a bit with the Stern+BAO case. Suppose for Linear
parametrization, here also the curves are open downwards in $\left(w_0
, w_1\right)$ space. The slope is sufficiently negative. Best fit is
situated at the third quadrant. All over a negative EoS is
indicated. Though for CPL and JBP it may be concluded that the
models in BD cosmology will not mind if we make our $w_1$ positive
and with a high magnitude. But of course these will be very
particular cases or exceptions which may not be physical.
%%%%%%%%%%%%%%%%%%%%%%%%%%%%%%%%%%%%%%%%%%%%%%%%%%%%%%%%%%%%%%%%%%%%%%%%%%%%%%%%%%%%%%%%%%%%%%%%%%
\section{Redshift-Magnitude Observations from Supernovae Type Ia}
\label{Redshift-Magnitude Observations from Supernovae Type Ia}
%%%%%%%%%%%%%%%%%%%%%%%%%%%%%%%%%%%%%%%%%%%%%%%%%%%%%%%%%%%%%%%%%%%%%%%%%%%%%%%%%%%%%%%%%%%%%%%%%%

The Supernova Type Ia experiments provided the main evidence for
the existence of DE. Since 1995, two teams of High-$z$ Supernova
Search and the Supernova Cosmology Project have discovered several
type Ia supernovas at the high redshifts
\cite{Perlmutter1,Riess1}. The observations directly measure the
distance modulus of a Supernovae and its redshift $z$
\cite{Kowalaski}. Now, take recent observational data,
including SNe Ia which consists of 557 data points and belongs to
the Union2 sample \cite{Amanullah}.

From the observations, the luminosity distance $d_{L}(z)$
determines the dark energy density and is defined by
\begin{equation}
d_{L}(z)=(1+z)H_{0}\int_{0}^{z}\frac{dz'}{H(z')}
\end{equation}
the apparent magnitude $m$ of a supernova and its redshift $z$ are
directly measured from the observations. The apparent magnitude
$\mu$ (the distance modulus - distance between absolute and
apparent luminosity of a distance object- for Supernovae) is
related to the luminosity distance $d_L$ of the supernova by the
relation:
\begin{equation}
\mu(z)=5\log_{10} \left[\frac{d_{L}(z)/H_{0}}{1~MPc}\right]+25
\end{equation}
The best fit of distance modulus as a function $\mu(z)$ of
redshift $z$ for our theoretical model and the Supernova Type Ia
Union2 sample are drawn in figure 4a, 4b and 4c. It is very clear
that for low redshifts $z<0.4$ CPL and JBP are
efficient enough to explain the Observational data (in the
background of the BD cosmology). The linear parametrization case is efficient enough upto $z=0.2$. Then it is over determined and after $z=0.6$ it is under determined.
\begin{figure}
~~~~~~~~~~~~~~~~~~~~~~~Fig.4a~~~~~~~~~~~~~~~~~~~~~~~~~~~~~~~~~~~~~~~
Fig.4b~~~~~~~~~~~~~~~~~~~~~~~~~~~~~~~~~~~~~~~Fig.4c\\
\includegraphics[height=2in, width=2in]{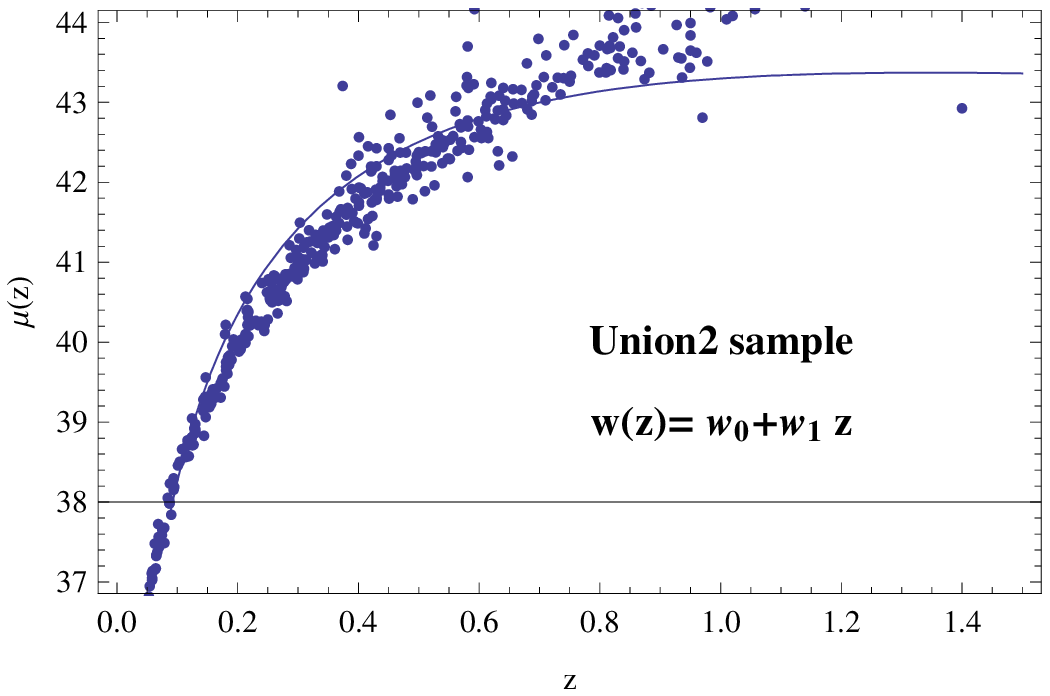}~~~~~~~\includegraphics[height=2in, width=2in]{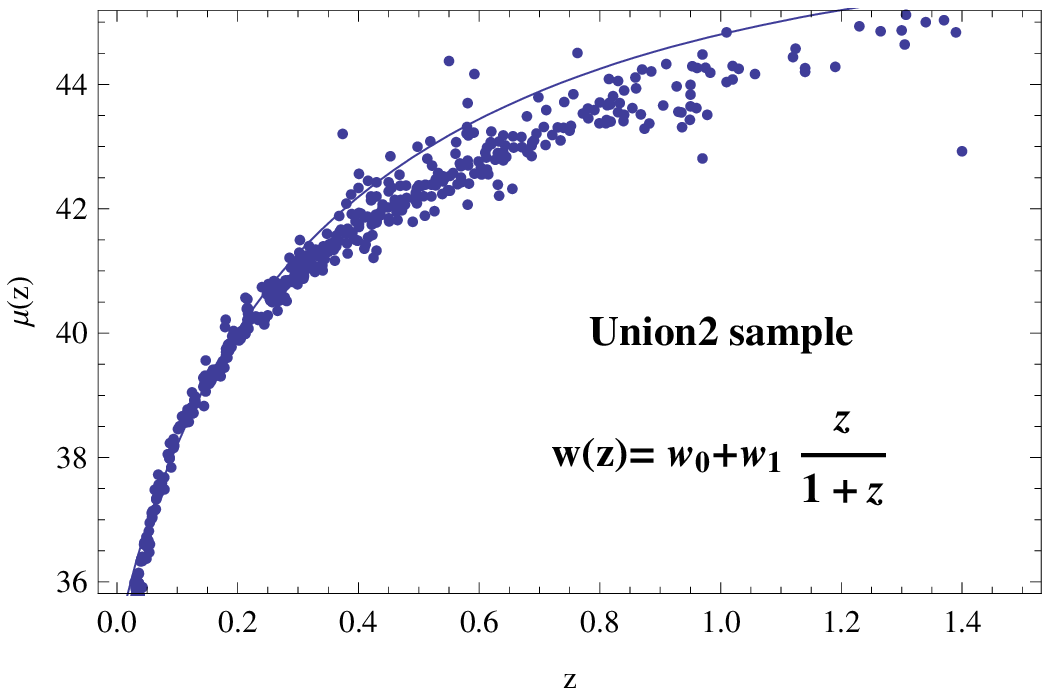}~~~~~~~
\includegraphics[height=2in, width=2in]{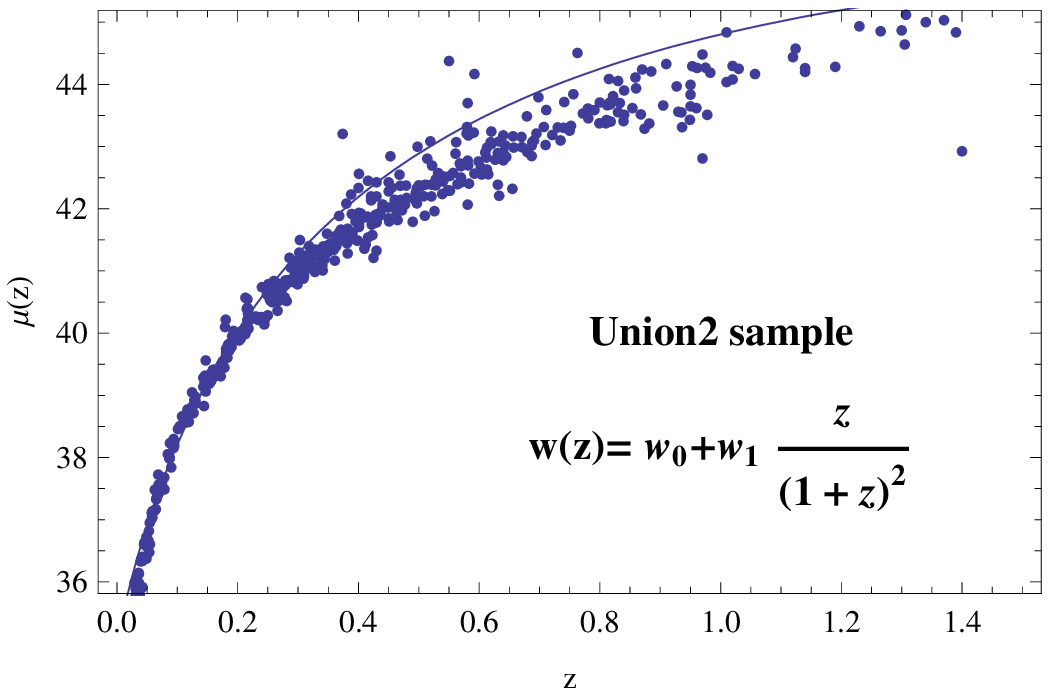}\\

\vspace{1cm}

Figs. 4(a), (b), (c) show the variation of $\mu (z)$ with $z$ for
Linear, CPL and JBP parameterizations respectively (Solid lines).
The dots denote the Union Sample. \vspace{1cm}
\end{figure}

%%%%%%%%%%%%%%%%%%%%%%%%%%%%%%%%%%%%%%%%%%%%%%%%%%%%%%%%%%%%%%%%%%%%%%%%%%%%%%%%%%%%%%%%%%%%%%%%%%%%%%%%%%%%%%%%%%%%%

 %%%%%%%%%%%%%%%%%%%%%%%%%%%%%%%%%%%%%%%%%%%%%%%%%%%%%%%%%%%%%%%%%%%%%%%%%%%%%%%%%%%%%%%%%%%%%%%%%%%%%%%%
\section{Brief Summary}\label{Summary}
%%%%%%%%%%%%%%%%%%%%%%%%%%%%%%%%%%%%%%%%%%%%%%%%%%%%%%%%%%%%%%%%%%%%%%%%%%%%%%%%%%%%%%%%%%%%%%%%%%%%%%%%

Though, this is the time to describe the outcomes of this work in
brief, we must say some important results of existing literature.
Fabris et al \cite{Fabris} have predicted that the consideration
of $\Lambda$CDM model might give the lowest $\chi^2$. It is true
that their main motive was to determine $H_0$, and ultimately, the
value of which was determined around 0.6 which quite resembles
with the pre-predicted values of $H_0$ ($=0.72 \pm 0.05$)
\cite{Spergel1}. For Brans-Dicke (BD) cosmology, they have
speculated the best value of $\omega_{BD}$ to be $-1.5$
(remarkably, this is conformally equivalent to General
Relativity). In this work, we determine the value of this
parameter from the density factors of different components of the
universe and few distinct parameters of Brans Dicke cosmology
itself. Using this value of $\omega_{BD}(=86720.9)$ our main
concern is to set the unknown parameters of the different DE
models' EoS parameter. So far, we have found the best fit
values of two unknown parameters $w_0$ and $w_1$ (in redshift
parametrization of DE) in the background of BD cosmology. One
important point to be signed is that previous study indicate that
structures can form in the Brans-Dicke model considered here
during all the evolution of the universe, after the radiative
phase, even the gravitational coupling is, at large scale,
repulsive. We have however fixed the $H_0$ at the beginning and
wanted to find out the confidence intervals of those parameters
which explicitly determine the nature of DE/ the DE EoS. Most
strange thing is always we got a open confidence contour (66\%,
90\%, 99\%). To do this we take twelve point red-shift vs Hubble's
parameter data and perform $\chi^2$ test. We present the
observational data analysis mechanism for Stern, Stern+BAO and
Stern+BAO+CMB observations. Minimizing $\chi^2$, we obtain the
best fit values of DE redshift parametrization parameters and draw
different confidence contours. Though the best fit values, found
out from different analysis pointed out towards a negative EoS at
$z=0$. Mathematically, it was showing a large range (actually
unbounded) ordered pair of $(w_0~, ~w_1)$ is allowed to stay in
the $1\sigma$ confidence contour. Hypothetically, it will not mind
if we keep our parameter at any place in that range. We will be
still in the confidence level. But, the values taken by the
parameters on their own, i.e., the physical values of them are not
forced to stay anywhere at that range. Rather, they will chose
their own positions! Preferably that will be inside the specified
zone. While giving the confidence range, BD is giving enough
liberty. As a modification of Einstein gravity it is very
interesting nature to follow. Finally, we examine the best fit of
distance modulus for our theoretical models and the Supernova Type
Ia Union2 sample and we found that for low redshifts $z<0.4$, all
the parameterizations are efficient enough to explain the
Observational data in the BD cosmology.\\

The concluding lines in a nutshell should be: In BD theory,
while all the observational data supported values of Hubble's
parameter, dark energy, dark matter, radiation densities etc have
been considered in an closed universe with a potential
proportional to the square of the scalar field present inside it,
if the scalar field is proportional to the square root of the
scale factor of the universal expansion, we get almost negative
EoS-s for the fluid present inside the universe (for three
particular fluids with redshift parametrizations of their EoS).
Unlike the other gravity theory results we do not get a closed
$1\sigma$ confidence contour for the parametric values
(considering the redshift parametrization). We get a open curve,
tendency of which says it is preferable to get negative parametric
values which ultimately would evolve negative EoS strictly
indicating the negative pressure inside the BD universe. Inclusion
of a scalar field has such an impression upon the inside-fluid's
EoS that in spite of being confined inside a short region,
infinite values of EoS are likely to have.\\

\begin{contribution}\\\\
{\bf Acknowledgement :}\\

UD thanks to CSIR, Govt. of India for providing research project
grant (No. 03(1206)/12/EMR-II) and RB also thanks to above CSIR
project for awarding Research Associate fellowship.

\end{contribution}


\frenchspacing
\begin{thebibliography}{100}
%%%%%%%%%%%%%%%%%%%%%%%%%%%%%%%%%%%BD Theory%%%%%%%%%%%%%%%%%%%%%%%%%%%%%%%%%%%%%%
\bibitem{Brans1} Brans, C., Dicke, R., H. : {\it Phys. Rev} {\bf 124} 925(1961).
\bibitem{Barrow1} Barrow, J. D., Maeda, K. : {\it Nucl Phys. B.} {\bf 341} 294(1990).
\bibitem{Reasenberg1} Reasenberg, R. D. et al : {\it ApJ} {\bf 234} L219(1979).
\bibitem{Bertotti1} Bertotti, B., Iess, L., Tortora, P:- {\it Nature} {\bf 425}, 374(2003).
\bibitem{Eddington1}  Eddington, A.S. :- {\it The mathematical Theory of Relativity}, {\bf Cambridge University Press} (1922);
Nordtvedt, K. :- {\it Phys. Rev. } {\bf 169}, 1017(1968);  Will, C.M.,  Nordtvedt, K. :- {\it Astrophys. J. } {\bf 177}
757 (1972);  Will, C.M.:- {\it Theory and Experiment in Gravitational Physics}, {\bf Cambridge Univ.
Press} (1993).
\bibitem{Will1} Will, C.M.:- {\it  Living Rev. Rel.} {\bf 4}, 4 (2001), gr-qc/0103036.
\bibitem{Shapiro1} Shapiro, I.I. :- {\it in General Relativity and Gravitation} {\bf 12}, edited by N. Ashby, D.F. Bartlett, and
W. Wyss, Cambridge University Press (1990), p. 313.
\bibitem{Williams}  Williams, J.G., Newhall, X.X., Dickey,  J.O. :- {\it Phys. Rev. D} {\bf  53}, 6730 (1996).
\bibitem{Perivolaropoulos} Perivolaropoulos, L. :- {\it Phys.Rev.D} {\bf 81}, 047501(2010).
\bibitem{Moon} Moon, T., Oh, P.:- arXiv 1302.3061v1.
\bibitem{Eubanks} Eubanks,T.M.,  Martin, J.O.,  Archinal, B.A.et al.:- {\it Bull. Am. Phys. Soc.}, Abstract No. K 11.05(1997), unpublished; draft at $ftp://casa.usno.navy.mil/navnet/postscript/prd_15.ps$
(1999); Shapiro,  S.S., Davis, J.L., Lebach, D. E.,  Gregory, J.S.:- {\it Phys. Rev. Lett.}  {\bf 92}, 121101(2004).
%%%%%%%%%%%%%%%%%%%%%%%%%%%%%%%%%%%DE%%%%%%%%%%%%%%%%%%%%%%%%%%%%%%%%%%%%
\bibitem{Riess1} Riess, A. G. et al :-
{\bf [Supernova Search Team Collaboration]}, {\it Astron. J.} {\bf 116}, 1009(1998)[{\it arXiv}:{\bf 9805201}(astro-ph)].
\bibitem{Perlmutter1} Perlmutter, S. et al :-
{\bf [Supernova Cosmology Project Collaboration]}, {\it ApJ} {\bf
517}, 565(1999)[{\it arXiv}:{\bf 9812133}(astro-ph)].
\bibitem{Spergel1} Spergel, D. N. et. al.: {\it Astrophys. J. Suppl.} {\bf 148} 175(2003).
\bibitem{Knop1}  Knop, R. A. et. al.: {\it Astrophys.J.} {\bf 598} 102(2003).
%%%%%%%%%%%%%%%%%%%%%%%%%%%%%%%%%%%%%%%%%%%%%%%%%%%%%%%%%%%%%%%%%%%%%%%%%
\bibitem{Melchiorri1}Melchiorri, A., Mersini, L., Trodden, M. :- {\it Phys. Rev. D} {\bf 68} 043509(2003).
\bibitem{Seljak1}Seljak, U., Slosar, A., McDonald, P.:- {\it JCAP} {\bf 0610} 014 (2006).
\bibitem{Cooray1} Cooray, A. R., Huterer, D. :- {\it Astrophys. J.} {\bf 513} L95(1999).
\bibitem{Upadhye1} Upadhye, A., Ishak, M., Steinhardt, P. :- {\it Phys Rev D} {\bf 72} 063501(2005).
\bibitem{Chevallier1} Chevallier, M., Polarski, D. :- {\it Int. J. Mod. Phys. D} {\bf 10} 213(2001).
\bibitem{Linder1}Linder, E. V. :- {\it Phys. Rev. Lett.} {\bf 90} 091301(2003).
\bibitem{Linden1}Linden, S., Virey, J. -M. :- {\it Phys. Rev. D} {78}
023526(2008).[{\it arXiv}:{\bf 0804.0389}].\\
\bibitem{Jassal1}Jassal, H. K., Bagla, J. S., Padmanabhan, T. :- {\it MNRAS} {\bf 356} L11(2005).
%%%%%%%%%%%%%%%%%%%%%%%%%%%%%%%%%%%%%%%%%%%%%%%%%%%%%%%%%%%%%%%%%%%%%%%%%%%%%%%%%%%
\bibitem{Errahmani1} Errahmani, A ; Ouali, T :- {\it Phys. Lett. B} {\bf 641} 357(2006).
 \bibitem{Kim1}Kim, H., Lee,  H. W., Myung, Y. S. :- {\it Phys.Lett. B} {\bf 628} 11(2005).
\bibitem{Arik1}Arik, M., Calik, M.C.:- {\it Mod.Phys.Lett.A} {\bf 21}1241(2006).
\bibitem{Kim2} Kim, H. :- {\it Mon.Not.Roy.Astron.Soc.} {\bf 364} 813(2005).
%%%%%%%%%%%%%%%%%%%%%%%%%%%%%%
\bibitem{Wu1} Wu, F.-Q., Chen, X. :- {\it PRD} {\bf 82}, 083003 (2010).
\bibitem{Li1}Li, Y.-C., Wu, F.-Q., Chen, X.:-   arXiv:1305.0055 [astro-ph.CO].
\bibitem{Fabris}  Fabris, J.C., Goncalves, S.V.B.,  Ribeiro, R. de Sa :- {\it   Grav.Cosmol.} {\bf 12} 49(2006).
%%%%%%%%%%%%%%%%%%%%%%%%%%%%%%%%%%%%%%%%%%%%%%%%%%%%%%%%%%%%%%%%%%%%%%%%%
\bibitem{Sen1} Sen, S.,Sen, A. A. :- {\it Phys. Rev. D} {\bf  63} 124006(2001); Sen, S., Seshadri, T. R. :- {\it Int. J. Mod. Phys. D}  {\bf 12} 445 (2003); Chakraborty, W., Debnath, U. :- {\it Int. J. Theor. Phys.} {\bf 48} 232 (2009);  Faraoni, V. :- {\it Phys. Rev. D} {\bf 62} 023504 (2000); Saini, T. D.,Raychauchaudhury,  S.,
Sahni, V.,Starobinsky,  A. A. :- {\it Phys. Rev. Lett.} {\bf 85} 1162 (2000).
%%%%%%%%%%%%%%%%%%%%%%%%%%%%%%%%%%%%%%%%%%%%%%%%%%%%%%%%%%%%%%%%%%%%%%%%%%%%%%%%
\bibitem{Stern} Stern, D. et al, 2010, JCAP 1002, 008.
\bibitem{Wu} Wu, P. and Yu, H., 2007, Phys. Lett. B 644, 16.
\bibitem{Paul1} Thakur, P., Ghose, S. and Paul, B. C., 2009, Mon. Not. R. Astron. Soc. 397, 1935.
\bibitem{Paul2} Paul, B. C., Ghose, S. and Thakur, P., arXiv:1101.1360v1
  [astro-ph.CO].
\bibitem{Paul3} Paul, B. C., Thakur, P. and Ghose, S., arXiv:1004.4256v1
 [astro-ph.CO].
\bibitem{Paul4} Ghose, S., Thakur, P. and Paul, B. C., arXiv:1105.3303v1
   [astro-ph.CO].
\bibitem{Chak} S. Chakraborty, U. Debnath and C. Ranjit, Eur. Phys. J. C, 72
2101 (2012).
\bibitem{Einstein} Eisenstein, D. J. et. al. :- {\it Astrophys. J.} {\bf 633}, 560(2005).
\bibitem{Bond} Bond, J. R. et. al.:- 1997, Mon. Not. Roy. Astron. Soc. 291, L33.
\bibitem{Efstathiou} Efstathiou, G., Bond, J. R. :-{\it MNRAS} {\bf 304}, 75(1999).
\bibitem{Nessaeris} Nessaeris, S., Perivolaropoulos, L.:- {\it JCAP} {\bf 0701}, 018(2007).
\bibitem{Komatsu1} Komatsu, E. et al: {\it Astrophys. J. Suppl.} {\bf 192}, 18(2011).
\bibitem{Kowalaski} Kowalaski et. al. :- {\it Astrophys. J.} {\bf 686}, 749(2008).
\bibitem{Alberto1} Alberto Vazquez, J., Bridges, M.,   Hobson, M.P.,  Lasenby,
A.N. :- [{\it arXiv}:{\bf 1205.0847}[astro-ph.CO]].
\bibitem{Amanullah} Amanullah, R. et al:- {\it Astrophys. J.} {\bf 716}, 712(2010).

\end{thebibliography}
\end{document}